\documentclass[11pt,a4paper]{article}

\usepackage{jheppub} 

\usepackage{multirow}
\usepackage{amsmath}
\usepackage{pifont}
\usepackage{float}
\usepackage{cancel}

\usepackage{epstopdf}

\allowdisplaybreaks

\usepackage{xcolor}

\usepackage{lineno} 

\usepackage{hyperref} 
\hypersetup{
    colorlinks,
    citecolor=black,
    filecolor=black,
    linkcolor=black,
    urlcolor=black
}

\RequirePackage{flushend}
\RequirePackage[numbers,sort&compress]{natbib}

\usepackage{slashed}

\newcommand{\dhalf}{\frac{d}{2}}

\newcommand{\Ds}{\displaystyle}

\newcommand{\nn}{\nonumber}

\newcommand{\Li}{\text{Li}}

\renewcommand{\(}{\left(}
\renewcommand{\)}{\right)}

\def\II{\hbox{{1}\kern-.25em\hbox{l}}}

\title{QCD factorization for chiral-odd parton quasi- and pseudo-distributions}

\author[a]{Vladimir~M.~Braun,}
\author[b]{Yao Ji,}
\author[a]{and Alexey~Vladimirov}

\subheader{
\begin{flushright}
{\small
{SI-HEP-2021-21, P3H-21-054}
}
\end{flushright}
}

\affiliation[a]{Institut f\"ur Theoretische Physik, Universit\"at Regensburg, D-93040 Regensburg, Germany}
\affiliation[b]{Theoretische Physik 1, Naturwissenschaftlich-Technische Fakult{\"a}t,
Universit{\"a}t Siegen, 57068 Siegen,\\ Germany}

\emailAdd{vladimir.braun@ur.de}
\emailAdd{yao.ji@uni-siegen.de}
\emailAdd{alexey.vladimirov@ur.de}

\abstract{
We study chiral-odd quark-antiquark correlation functions suitable for lattice calculations 
of twist-three nucleon parton distribution functions $h_L(x)$ and $e(x)$, and also the twist-two transversity distribution $\delta q(x)$.  
The corresponding factorized expressions are derived in terms of the twist-two and twist-three collinear distributions to one-loop accuracy.
The results are presented both in position space, as the factorization theorem for Ioffe-time distributions,
and in momentum space, for quasi- and pseudo-distributions. We demonstrate that the twist-two part of the $h_L$ quasi(pseudo)-distribution
can be separated from the twist-three part by virtue of an exact Jaffe-Ji-like relation.
}

\begin{document} 

\maketitle 

\section{Introduction}

Parton distribution functions (PDFs) are the quintessential component of the modern picture of the hadron structure. The twist-two PDFs have a probabilistic interpretation as parton density distributions within hadrons and give rise to the leading contribution to the majority of observables. In contrast, the twist-three distributions are related to the quantum-mechanical interference between a single parton and a gluon-parton pair. Their contributions are usually suppressed by the hard scale, although they can be enhanced in certain kinematic regions. Nevertheless, twist-three effects are vital for the understanding of hadron structure at a quantum level. 

Quantifying twist-three effects is a challenging task. The lattice QCD has the potential to explore the twist-three distributions by a calculation of specially designed Euclidean observables. In particular, the nucleon matrix element of the local twist-three chiral-even operator of the lowest dimension was first calculated on the lattice in Ref.~\cite{Gockeler:2005vw}, and such studies will undoubtedly continue. The major difficulty of this approach is that due to the loss of continuous space-time symmetry, the twist-three operators on the lattice mix with the leading-twist operators of lower dimension so that the renormalization procedure becomes highly nontrivial. 

In recent years there has been increasing interest in the possibility to determine PDFs from custom-made Euclidean correlation functions, bypassing Wilson's operator product expansion. The general scheme of such calculations is to consider a product of suitable local currents at a space-like separation and match the lattice calculation of this quantity to the perturbative expansion in terms of collinear distributions. This can be done both in position and momentum spaces. Several proposals to implement this idea exist \cite{Aglietti:1998ur,Detmold:2005gg,Braun:2007wv,Ji:2013dva,Ma:2014jla,Chambers:2017dov}. The particular choice of the correlation function of the quark and antiquark fields connected by the Wilson line~\cite{Ji:2013dva} received the most attention, see, e.g., \cite{Ji:2020ect,Constantinou:2020pek} for a review. In order to distinguish these objects from their light-cone analogs, one usually refers to such space-like correlation functions as pseudo-PDFs (e.g.~\cite{DelDebbio:2020rgv})  or quasi-PDFs (e.g.~\cite{Alexandrou:2020qtt}), depending on the actual implementation. 

The first lattice simulations of twist-three PDFs within the qPDF approach appeared recently~\cite{Bhattacharya:2020cen,Bhattacharya:2021moj}, demonstrating the feasibility of such studies. The discussion of the collinear limit of the twist-three qPPFs and pPDFs presented in~\cite{Bhattacharya:2020xlt,Bhattacharya:2020jfj} is, however, not complete. In Ref.~\cite{Braun:2021aon} we have formulated the factorization theorem for the axial-vector qPDF (and pPDF) to twist-three accuracy, which is more involved compared to the twist-two case~\cite{Izubuchi:2018srq} and calculated the corresponding coefficient functions to one-loop accuracy. In this paper, we repeat the same analysis for the chiral-odd case and formulate the factorization theorem for the quasi(pseudo) distributions related to the twist-three PDFs $h_L(x)$ and $e(x)$. As a byproduct of this calculation, we also derive the one-loop coefficient function for the twist-two transversity PDF $\delta q(x)$, for which, it seems, there is no agreement in the literature (cf. \cite{Liu:2018hxv,Alexandrou:2018eet}).
 
This direction of studies is interesting as the chiral-odd twist-three PDFs $e(x)$ and $h_L(x)$ are poorly known. They show up in observables that involve other chiral-odd functions and thus are challenging to extract from experimental data. As a matter of principle, these functions can be directly extracted from the single- and double-spin asymmetries in semi-inclusive deep inelastic scattering (SIDIS) and Drell-Yan process \cite{Tangerman:1994bb,Bacchetta:2003vn,Courtoy:2014ixa,Koike:2008du}. This is, however, problematic due to a lack of very precise data. Alternatively, twist-three distributions can be accessed in the small-$b$ limit of transverse momentum distributions (TMDs). This approach looks more promising as the relevant observables are easier to measure, and hence the data is of better quality. For example, the Qiu-Sterman function $T$ has been extracted in Refs.~\cite{Bury:2020vhj,Bury:2021sue}  from the analysis of the small-$q_T$ part of the transverse spin asymmetry data. Chiral-odd distributions can therefore be addressed similarly by studying the chiral-odd TMDs, see \cite{Kanazawa:2015ajw,Scimemi:2018mmi,Moos:2020wvd} for the relevant relations. In this way, the Qiu-Sterman projection of the underlying quark-gluon correlation function that is related to $e(x)$ gives rise to the Boer-Mulders function and, thus, could be observed in the $\cos(2\phi)$-modulation in unpolarized Drell-Yan processes~\cite{Boer:1997nt}. In a different context, the role of the twist-three PDF $e(x)$ in the proton mass decomposition has also been discussed in~\cite{Hatta:2020iin,Ji:2020baz}.

In Ref.~\cite{Braun:2021aon} we have pointed out that the extraction of the chiral-even twist-three PDF $g_T(x)$ from lattice calculations is complicated by the breakdown of the Wandzura-Wilczek relation at the qPDF (pPDF) level. Thus, the separation of twist-three contributions of interest from the dominant twist-two terms at the level of lattice data appears to be highly nontrivial. For chiral-odd distributions, the situation is better as the qPDF (pPDF) related to $e(x)$ does not contain any twist-two admixture, and for $h_L(x)$ we are able to derive an exact Jaffe-Ji-like relation for the twist-two part. 

This work follows Ref.~\cite{Braun:2021aon} both conceptually and methodologically, so we present only the necessary definitions and the final expressions for the factorization theorems and NLO coefficient functions and omit technical details. Sec.~\ref{sec:1} is introductory, and it contains general definitions and notation. The tree-level factorization expressions in terms of quark-gluon correlation functions, QCD equation of motion (EOM) relations, and PDF definitions are given in Sec.~\ref{sec:tree}. The Sec.~\ref{sec:nlo} is devoted to the calculation of the NLO corrections and contains our main results. The position-space expressions are given in Sec.~\ref{sec:qITDs}, and the  momentum space ones are collected in Sec.~\ref{sec:pPDFs} and  Sec.~\ref{sec:qPDFs} for the pPDFs and qPDFs, respectively. The concluding Sec.~\ref{sec:disc} contains the summary and a short discussion. Two appendices supplement the main text: App.~\ref{app:evolution_kernel} contains the evolution kernel for the twist-three chiral-odd distributions, and in App.~\ref{app:OPE} the operator-level results are presented for the first order expansion in $\alpha_s$ of the chiral-odd quark-antiquark correlation functions in position space.

\section{Preliminaries}
\label{sec:1}

We consider the nucleon matrix elements of a product of 
quark and antiquark fields for chiral-odd Dirac structures in the forward limit,
\begin{align}
& \langle p,s|\mathcal{O}^{\Gamma}(z,0)|p,s\rangle\,,
\label{NME}
\end{align}
where
\begin{align}\label{def:op-main}
\mathcal{O}^{\Gamma}(z,0)& =\text{T}\big\{\bar q(z)\Gamma [z,0]q(0)\big\},
\end{align}
and $\Gamma = \II $ or $\Gamma = i\sigma_{\mu\nu}\gamma^5 = - \tfrac12 \epsilon_{\mu\nu\alpha\beta}\sigma^{\alpha\beta}$ (in four dimensions).
Here  $z^\mu$ is a four-vector, $|p,s\rangle$ stands for the nucleon state with momentum $p$ and spin $s$, 
 $q$ and $\bar q$ denote the quark and antiquark fields, respectively, and $[z,0]$ is the straight Wilson line 
in the fundamental representation of the gauge group rendering the operator gauge invariant,
\begin{eqnarray}
[z,0]=P\exp\Big(ig\int_{0}^{1}\!d\sigma\, z^\mu A_\mu(\sigma z)\Big).
\end{eqnarray}
We follow the convention   
$\gamma^5=i\gamma^0\gamma^1\gamma^2\gamma^3$ and $\sigma^{\mu\nu}=\tfrac{i}{2}(\gamma^\mu\gamma^\nu-\gamma^\nu\gamma^\mu)$.
The flavor indices of the quark fields are omitted for brevity.
For space-like separations $z^2<0$, which are of interest in the context of lattice calculations in Euclidean space, 
the time-ordering in Eq.~\eqref{def:op-main} is in fact redundant. 
In this work, we calculate the matrix elements \eqref{NME} at the next-to-leading order (NLO) and twist-three accuracy in QCD
taking the collinear parton distributions as the nonperturbative input.
Note that in the forward kinematics considered here, the matrix elements
only depend on the distance $z$ between the fields.
Thus, without loss of generality, we keep the quark field  $q$ at the origin.

We tacitly assume that the nonlocal operator \eqref{def:op-main} is renormalized,
\begin{eqnarray}\label{def:ZZO}
\mathcal{O}^{\Gamma}(z,0)&=& Z_{\mathcal O} {\cal O}_{\text{bare}}^{\Gamma}(z,0)\,.
\end{eqnarray}
The renormalization constant $Z_{\mathcal O}$ in the $\overline{\text{MS}}$ scheme 
is known to three-loop accuracy~\cite{Chetyrkin:2003vi,Braun:2020ymy}.
It is the same for all Dirac structures. 
Both sides of Eq.~\eqref{def:ZZO} depend on the renormalization scale $\mu_R$. 
Here and in the following, we do not indicate the dependence on $\mu_R$, unless it is important for understanding. 

In  renormalization  schemes  with  an  explicit  regularization  scale,  the Wilson  line  in  Eq.~\eqref{def:op-main}  
suffers  from  an  additional  linear  ultraviolet  divergence \cite{Dotsenko:1979wb} which  has  to  be  removed.   
This  can  be  done  by introducing a residual mass term that absorbs such divergences,  as in the case of heavy quark effective theory, 
or, alternatively, by forming a suitable  ratio of matrix elements involving the same 
operator \cite{Orginos:2017kos,Braun:2018brg}. This issue has been discussed extensively in the literature 
so that in our opinion no further explanation is needed.

The nucleon matrix element \eqref{NME} can be parametrized in terms of several 
invariant functions. We define 
\begin{align}\nn
\langle p,s|\mathcal{O}^{i\sigma^{\mu z} \gamma^5}(z,0)|p,s\rangle=&2 s_T^\mu (p\cdot z) \mathcal{H}_1(p\cdot z,z^2)
-2\lambda_z\(z^\mu-\frac{z^2}{(p\cdot z)}p^\mu\) M\mathcal{H}_L(p\cdot z,z^2),
\\\label{def:ITD}
\langle p,s|\mathcal{O}^{\scriptsize\II}(z,0)|p,s\rangle=&2M \mathcal{E}(p\cdot z,z^2).
\end{align}
Here  $\sigma^{\mu z}=\sigma^{\mu\nu}z_\nu$, $M^2=p^2$ is the nucleon mass squared, 
$s^\mu$ is the spin vector normalized as $s^2=-1$, $(s\cdot p)=0$ for transversely polarized nucleon state, and the dimensionless parameter
\begin{eqnarray}
\lambda_z= M\frac{(s\cdot z)}{(p\cdot z)}.
\end{eqnarray}
The subscript $T$ indicates the projection of the nucleon spin vector onto the transverse plane, 
which is orthogonal to both  $z^\mu$ and $p^\mu$:
\begin{align}
s_T^\mu = g_T^{\mu\nu} s_\nu\,,
\end{align} 
with
\begin{eqnarray}\label{def:gTtensor}
g_T^{\mu\nu}=g^{\mu\nu}-(z^\mu p^\nu+p^\mu z^\nu)\frac{p_z}{p_z^2-z^2 p^2}+p^\mu p^\nu \frac{z^2}{p_z^2-z^2p^2}+z^\mu z^\nu \frac{p^2}{p_z^2-z^2 p^2}\, .
\end{eqnarray}

We refer to the variable 
\begin{align}
\zeta =  p_z = (p\cdot z)
\end{align}
as  the Ioffe time~\cite{Ioffe:1969kf}. 
The similarly defined (regularized) invariant functions for the 
light-like separations $z^2=0$ give rise to the chiral-odd parton distributions in position space (see below) are 
called Ioffe-time distributions (ITDs)~\cite{Ioffe:1969kf,Braun:1994jq,Orginos:2017kos}.
The terminology for the corresponding off-light cone position-space matrix elements is not well established: 
they are referred to as  quasi-ITDs in~\cite{Braun:2018brg,Braun:2021aon} 
and pseudo-ITDs in~\cite{Radyushkin:2018cvn,Joo:2019jct,Joo:2020spy}. In this work, we use the abbreviation 
qITD for the functions $\mathcal{E}(\zeta,z^2)$, $\mathcal{H}_1(\zeta,z^2)$ and
 $\mathcal{H}_L(\zeta,z^2)$. 

The three qITDs $\mathcal{E}_1$, $\mathcal{H}_1$ and $\mathcal{H}_L$ at small $z^2$ match
collinear parton distributions (PDFs). 
QCD factorization for $\mathcal{E}$ and $\mathcal{H}_L$ is more complicated as compared to  $\mathcal{H}_1$, as both of them 
recieve contributions of twist-three quark-antiquark-gluon collinear distributions. In this paper we 
formulate the factorization theorem for $\mathcal{E}$ and $\mathcal{H}_L$ and calculate the necessary coefficient functions to one-loop accuracy.  

Our method of calculation is based on the light-ray operator product expansion (OPE) \cite{Anikin:1978tj,Balitsky:1987bk} 
in combination with the background field technique~\cite{Abbott:1980hw,Abbott:1981ke}. 
We used the same approach for chiral-even distributions in our previous publication~\cite{Braun:2021aon}, where it is described in details. 
The present case only differs from Ref.~\cite{Braun:2021aon} by the Dirac structure $\Gamma$, so that the calculation
follows the same route. The actual computation is done in position space. 
The coefficient functions for qPDFs and pPDFs in momentum fraction space are then obtained 
by the appropriate Fourier transform, as defined in the secs.~\ref{sec:pPDFs} and \ref{sec:qPDFs}.

\section{Chiral-odd parton distributions and quark-antiquark-gluon correlation functions}
\label{sec:tree}

At the tree level, the operators \eqref{def:op-main} in the $z^2\to0$ limit can be identified with the 
corresponding light-ray operators, cf.~\cite{Braun:2021aon}, whose nucleon matrix elements 
define parton distribution functions. 
In this work, we need the following light-cone matrix elements
\begin{align}
\label{def:H1}
g^{\mu\nu}_{T}\langle p,s|\bar q(z)i\sigma_{\nu z}\gamma^5 q(0)|p,s\rangle &= 2  s^\mu_T \zeta \int_{-1}^1 \! dx\, e^{ix\zeta} \delta q(x)
\equiv 2 s^\mu_T \zeta\, \widehat {\delta q}(\zeta)\, ,
\\
\label{def:HL}
\langle p,s|\bar q(z)i\sigma^{p z}\gamma^5 q(0)|p,s\rangle &= -2M \lambda_z \zeta \int_{-1}^1 \! dx\, e^{ix\zeta} h_L(x)
\equiv -2M \lambda_z \zeta  \widehat{h}_L(\zeta)\, ,
\\
\label{def:e}
\langle p,s|\bar q(z)q(0)|p,s\rangle& = 2M \int_{-1}^1 \! dx\, e^{ix\zeta} e(x)\equiv 2M \widehat{e}(\zeta)\, .
\end{align}
Here and in what follows, we use the ``hat'' notation for the PDFs in position space (Ioffe-time distributions~\cite{Ioffe:1969kf,Braun:1994jq}). Thus, at the tree level (and neglecting terms $\mathcal{O}(z^2)$) we have
\begin{align}\label{tree-for-qITDs}
\mathcal{H}_1(\zeta,z^2)=\widehat{\delta q}(\zeta)\,,
\qquad
\mathcal{H}_L(\zeta,z^2) = \widehat h_L(\zeta) \,, 
\qquad  
\mathcal{E}(\zeta,z^2) = \widehat e(\zeta)\,.
\end{align}

The transversity PDF $\delta q(x)$ is purely twist two. For $x>0$ and $x<0$ it defines the distribution of the (transversely polarized) quarks and antiquarks (with a sign minus) in the nucleon, respectively. 

In contrast, the function $h_L(x)$ can be decomposed into twist-two and twist-three parts,
according to the (geometric) twist of the contributing operators: 
\begin{align}
h_L(x)=  h_L^{\text{tw2}}(x) + h_L^{\text{tw3}}(x)\,, && \widehat h_L(\zeta)=  \widehat h_L^{\text{tw2}}(\zeta) + \widehat h_L^{\text{tw3}}(\zeta)\,. 
\end{align}
One obtains for the twist-two part \cite{Jaffe:1991kp},
\begin{align}
\widehat{h}^{\text{tw2}}_L(\zeta)&=2\int_0^1 d\alpha\, \alpha\,\widehat{\delta q}(\alpha \zeta), 
\notag\\
{h}^{\text{tw2}}_L(x) &= 
2x\Big[\theta(x) \int_x^1\frac{dy}{y^2} \delta q(y)
- \theta(-x) \int_{-1}^x\frac{dy}{y^2} \delta q(y)\Big].
\label{hL-tw2}
\end{align}
A simple method to derive these relations is the following. 
The leading-twist contribution of the light-ray operator must satisfy the equation 
\cite{Balitsky:1989ry},
\begin{align}\label{tw2-def}
\frac{\partial}{\partial z^\mu}[\bar q(z)  \sigma_{\mu\nu} z^\nu i\gamma_5 q(0)]_{\text{tw2}} =0\,. 
\end{align}
Taking the nucleon matrix element of this operator identity and its parametrization in \eqref{def:ITD}, 
\eqref{def:H1}, \eqref{def:HL}, 
one obtains a first-order differential equation (see e.g.~\cite{Scimemi:2018mmi})
\begin{align}
\frac{d}{dx} \Big[x h^{\text{tw2}}_L(x)\Big] = 2 h^{\text{tw2}}_L(x)-  2 \delta q (x).  
\end{align}
It is easy to check that the expression in \eqref{hL-tw2} satisfies this equation.

Beyond the tree level, the PDFs become scale-dependent and the tree-level relations in~\eqref{tree-for-qITDs}  
become decorated by the perturbatively 
calculable coefficient functions. In addition, the twist-three quark-antiquark distributions
 $e$ and $h^{\text{tw3}}_L$ get mixed with the three-particle quark-antiquark-gluon 
correlation functions  giving rise to additional contributions that do not have a 
probabilistic interpretation. It is important to realize that  quark-antiquark and quark-antiquark-gluon contributions 
cannot be taken as independent degrees of freedom as they are related by QCD equations 
of motion (EOM). Neglecting operators with total derivatives that do not 
contribute to forward matrix elements, one obtains \cite{Braun:1989iv}
\begin{align}
\bar q(z)i\sigma^{\mu z}\gamma^5 q(0)&=[\bar q(z)i\sigma^{\mu z}\gamma^5 q(0)]_{\text{tw2}}
-\frac{z^\mu}{2}\int_0^1\! du\! \int_{0}^u \! dv \, u(2v\!-\!u)\, \bar q(uz)\,\sigma^{\nu z}\gamma^5 \, g F_{\nu z}(vz)q(0)\,,
\notag\\ 
\bar q(z)q(0)&=\bar q(0)q(0)+\frac{1}{2}\int_0^1\! du\! \int_{0}^u\! dv\, 
\bar q(uz)\,\sigma^{\mu z} \, g F_{\mu z}(vz)q(0)\, .
\label{EOM}
\end{align}
These operator relations are exact in QCD with massless quarks so that they must be satified by all matrix elements. 
In the first line, the first term on the r.h.s. gives rise to the twist-two distribution  $h^{\text{tw2}}_L$, and the second  
term provides a representation for $h^{\text{tw3}}_L$ in terms of a certain integral of the quark-antiquark-gluon correlation function
defined below. The second relation gives rise to a similar representation for $e(x)$ in terms of a (different)  
quark-antiquark-gluon correlation function, up to an additional local term.
It is convenient to consider the full set of quark-antiquark-gluon operators as the operator basis at the intermediate step.%
\footnote{Using an operator basis with the gluon strength tensor replaced by a transverse derivative is another option \cite{Ellis:1982cd}.
This choice usually leads to simpler coefficient functions at the cost of complicating the evolution equations.} 
Using \eqref{EOM}, a part of the resulting expressions can be
brought back to the two-particle form, but this reduction does not hold for the full result.

Nucleon matrix elements of the three-particle twist-three operators define position-space quark-antiquark-gluon 
correlation functions: 
\begin{align}
\langle p,s|\bar q(az)\,\sigma^{\mu z}\gamma^5 \, g F_{\mu z}(bz)q(cz) |p,s\rangle &= 4 \lambda_z \zeta^2 M \,\widehat{H}(a\zeta,b\zeta,c\zeta)\,.
\label{def:HH}
\\
\langle p,s|\bar q(az)\,\sigma^{\mu z} \, g F_{\mu z}(bz)q(cz) |p,s\rangle 
&= 4 \zeta^2 M \,\widehat{E}(a\zeta,b\zeta,c\zeta)\, .
\label{def:EE}
\end{align}
They are related to the correlation functions in  momentum fraction space by the Fourier transformation 
\begin{align}\label{def:twist-3-Fourier}
\widehat{H}(\zeta_1,\zeta_2,\zeta_3)& =\int [dx]\, e^{-i(\zeta_1x_1+\zeta_2x_2+\zeta_3x_3)}H(x_1,x_2,x_3),
\\
\widehat{E}(\zeta_1,\zeta_2,\zeta_3)& =\int [dx]\, e^{-i(\zeta_1x_1+\zeta_2x_2+\zeta_3x_3)}E(x_1,x_2,x_3),
\end{align}
where 
\begin{align}\label{def:[dx]}
\int [dx]&\equiv\int_{-1}^1 \!dx_1dx_2dx_3\,\delta(x_1+x_2+x_3)\,.
\end{align}
The functions $H$ and $E$ obey the discrete symmetry relation~\cite{Scimemi:2018mmi}
\begin{align}\label{def:H-sym}
H(x_1,x_2,x_3)&= -H(-x_3,-x_2,-x_1)\,,
\\\label{def:E-sym}
E(x_1,x_2,x_3)&= E(-x_3,-x_2,-x_1)\,.
\end{align}
As a consequence, for their position-space versions $\widehat{E}$ and $\widehat{H}$ the following relations hold,
\begin{eqnarray}\label{sym:integral-H}
\int_0^\alpha d\beta \,w(\beta)\widehat H(\alpha\zeta,\beta\zeta,0)&=&-\int_0^\alpha d\beta \,w(\alpha-\beta)\widehat H(\alpha\zeta,\beta\zeta,0),
\\\label{sym:integral-E}
\int_0^\alpha d\beta \,w(\beta)\widehat E(\alpha\zeta,\beta\zeta,0)&=&\int_0^\alpha d\beta \,w(\alpha-\beta)\widehat E(\alpha\zeta,\beta\zeta,0),
\end{eqnarray}
for any non-singular function $w(\alpha)$. These relations are essential to simplify the weight factors in the integrals, see Sec.~\ref{sec:nlo}.

The relation $x_1+x_2+x_3=0$ in \eqref{def:[dx]} imposed by the translational invariance in the forward kinematics implies that the twist-three PDFs are functions of two variables. 
However, the three-variable notation used here is more convenient for many reasons. First, it simplifies the symmetry 
relations \eqref{def:H-sym}, \eqref{def:E-sym}.  
Second, twist-three correlation functions have different parton interpretations for each kinematic 
domain $x_i\lessgtr 0$~\cite{Jaffe:1983hp} and can most naturally be presented using the three-component
barycentric coordinates, see~\cite{Braun:2009mi,Scimemi:2019gge}. 

With these definitions, we obtain  for the twist-three PDFs
\begin{align}
\widehat{h}^{\text{tw3}}_L(\zeta)&=\zeta^2 \int_0^1 d\alpha \int_{0}^\alpha d\beta \,\alpha\,(2\beta-\alpha )
\widehat{H}\(\alpha \zeta,\beta \zeta,0\),
\label{tree:h-position}
\\
\label{tree:e-position}
\widehat{e} (\zeta)&= \Sigma_{q} + \zeta^2 \int_0^1 d\alpha \int_{0}^\alpha d\beta \,\widehat{E}\(\alpha \zeta,\beta \zeta,0\),
\end{align}
where $\Sigma_{q}$ is the local contribution related to the 
nucleon $\sigma$-term,
\begin{align}\label{def:sigma-term}
\langle p,s|\bar q(0)q(0)|p,s\rangle = 2M\Sigma_{q}\,.
\end{align}
It is often convenient to separate the local contribution (\ref{def:sigma-term}) from the rest. 
In what follows we use the subscript ``nl'' (nonlocal) for quantities in which the sigma-term contribution is subtracted, e.g., 
\begin{align}
 \widehat{e} (\zeta)= \Sigma_{q} +  \widehat{e}_{\rm nl} (\zeta)\,, 
&& \widehat{e}_{\rm nl} (\zeta) =  \zeta^2 \int_0^1 d\alpha \int_{0}^\alpha d\beta \,\widehat{E}\(\alpha \zeta,\beta \zeta,0\)\,.
\end{align}

Going over to the momentum fraction space, one obtains
\begin{align}
h^{\text{tw3}}_L(x) &=2\int [dx]\int^1_0d\alpha\,\alpha\bigg(\frac{\delta(x+\alpha x_1)}{x_1x_2}-\frac{\delta(x-\alpha x_3)}{x_2x_3}-\frac{\delta(x+\alpha x_1)}{x_2^2}+\frac{\delta(x-\alpha x_3)}{x_2^2}\notag\\
&\hspace*{4cm}+\frac{\delta(x-x_3)}{x_2x_3}-\frac{\delta(x+x_1)}{x_1x_2}\bigg)H(x_1,x_2,x_3)\, ,
\\
e(x) &= \Sigma_q\delta(x) + \int [dx]  \left(\frac{\delta(x+x_1)}{x_1x_2}+\frac{\delta(x-x_3)}{x_2x_3}+\frac{\delta(x)}{x_1x_3}\right)E(x_1,x_2,x_3)\,.\label{tree:momentum}
\end{align}
In these expressions one can get rid of some integrations using the delta-functions, 
but the resulting expressions contain a multitude of integration regions, therefore we 
prefer to leave it in this form. 
We stress that these relations are exact in QCD.

The scale dependence of the correlation functions $E$ and $H$ is autonomous, 
in the sense that it does not involve other distributions. 
The evolution equations take the form
\begin{eqnarray}\label{th:evolution-gen}
\mu_F^2 \frac{d}{d\mu_F^2} H(x_1,x_2,x_3;\mu_F) &=& \int [dy]\, K_H(x,y;\alpha_s(\mu_F))\,H(y_1,y_2,y_3;\mu_F),
\\
\mu_F^2 \frac{d}{d\mu_F^2} E(x_1,x_2,x_3;\mu_F) &=& \int [dy]\, K_E(x,y;\alpha_s(\mu_F))\,E(y_1,y_2,y_3;\mu_F),
\end{eqnarray}
where the kernels $K_{E/H}$ are functions of six variables $x= \{x_1,x_2,x_3\}$ and $y = \{y_1,y_2,y_3\}$. 
The leading-order (LO) expressions of the evolution kernels can be found in \cite{Braun:2009mi,Ji:2014eta}. 
They are relatively compact in position space but rather lengthy in terms of the momentum fractions.
For completeness, we present the position-space kernels~\cite{Braun:2009mi} in App.~\ref{app:evolution_kernel}.

As a final remark, in Refs.~\cite{Scimemi:2018mmi,Moos:2020wvd}
a different notation for the chiral-odd twist-3 correlation functions is used, 
\begin{align}
\langle p,s|\bar q(az)\,i\sigma^{\mu z}\gamma^5\,g F^{\nu z}(bz)q(cz)|p,s\rangle
 &=2 \zeta^2 M\Big( \epsilon^{\mu\nu}_T \widehat{\delta T}_\epsilon(a\zeta,b\zeta,c\zeta)
+
i\lambda_z g_T^{\mu\nu} \widehat{\delta T}_g(a\zeta,b\zeta,c\zeta)\Big),
\end{align}
where $\epsilon^{\mu\nu}_T=p_\alpha z_\beta \epsilon^{\alpha \beta \mu\nu}/\zeta$ (and $z^2=0$).
The relation to our notation is $\delta T_g=H$ and $\delta T_\epsilon=E$.

\section{QCD factorization and one-loop results}
\label{sec:nlo}

Beyond the tree level, the expressions in \eqref{tree-for-qITDs} can be generalized to the following factorization theorems schematically:
\begin{align}
\mathcal{H}_1(\zeta,z^2) &= 
\int_0^1d\alpha\, C_1(\alpha, z^2\mu_F^2)\,\widehat{\delta q}(\alpha\zeta;\mu_F)\, +\ldots\,,
\\\nn
\mathcal{H}_L^{\text{tw2}}(\zeta,z^2) &= 
\int_0^1d\alpha\, C_L^{\text{tw2}}(\alpha, z^2\mu_F^2)\,\widehat{\delta q}(\alpha\zeta;\mu_F)\, +\ldots\,,
\\
\mathcal{H}_L^{\text{tw3}}(\zeta,z^2) &= 
\zeta^2\int_0^1 d\alpha\int_0^1 d\beta\, C_L^{\text{tw3}}(\alpha, \beta, z^2\mu_F^2)\widehat{H}(\alpha\zeta,\beta \zeta,0;\mu_F)
 +\ldots\,,
\notag\\
   \mathcal{E}(\zeta,z^2) &= C_\sigma (z^2\mu_F^2) \Sigma_q(\mu_F^2) +  
\zeta^2\int_0^1 d\alpha\int_0^1 d\beta\, C_S (\alpha, \beta, z^2\mu_F^2)\widehat{E}(\alpha\zeta,\beta \zeta,0;\mu_F)
 +\ldots\,,\notag
\end{align}
where the ellipses stand for terms $\mathcal{O}(z^2)$,
and the coefficient functions $C_k$ are given 
by a series expansion in the QCD coupling
\begin{align}
a_s  = \frac{\alpha_s(\mu_F)}{4\pi}\,.
\end{align}
We write
\begin{align}
 C_i = C_i^{(0)} + a_s\, C_i^{(1)} + \ldots  
\end{align}
with the tree-level expressions
\begin{eqnarray}\nn
C_1^{(0)}&=&\delta(1-\alpha)\,,\qquad C_h^{\text{tw2}(0)} = 2 \alpha\,, 
\\
C_h^{\text{tw3}(0)}&=& \alpha (2\beta-\alpha)\theta(\alpha-\beta)\,,
 \qquad C_e^{(0)} = \theta(\alpha-\beta)\,.
\end{eqnarray}
The validity of QCD factorization in this form is a direct consequence of the existence of the 
operator product expansion (OPE).
The factorization theorems for pseudo- and quasi-distributions in momentum space are obtained by 
the Fourier transform of the above expressions and do not require any additional justification. 
Explicit expressions  are given below.

The main purpose of this work is to compute the coefficient functions $C^{(1)}_{k}$. The calculation is done using the light-ray OPE 
and the background field technique, following the similar calculation for the chiral-even case in Ref.~\cite{Braun:2021aon}. 
Thus we present here only the final expressions for various distributions and omit technical details. 
The operator level expressions are collected in App.~\ref{app:OPE}.

As already discussed in Sec.~\ref{sec:tree}, 
the twist-three PDFs $\widehat h_L^{\text{tw3}}$ and $\widehat e$ are related to the more general three-particle distributions by 
Eqs.~\eqref{tree:h-position}, \eqref{tree:e-position}. Conversely, these identities allow one to rewrite a part of the three-particle contributions back 
to the two-particle form. This can be done as follows. Integrating Eqs.~\eqref{tree:h-position}, \eqref{tree:e-position} 
with an arbitrary weight function $f(\tau)$ one finds after simple algebra  
\begin{align}
\label{2pt->3pt:H}
\int^1_0\!d\tau\, f_h(\tau)\, h_L^{\text{tw3}}(\tau\zeta)&
=\zeta^2\int^1_0\!d\alpha\,
g_h(\alpha)
\int^\alpha_0 d\beta\, (2\beta-\alpha) \widehat H(\alpha\zeta,\beta\zeta,0)\,,
\\
\label{2pt->3pt:E}
\int^1_0\!d\tau\, f_e(\tau)\,e_{\rm nl}(\tau\zeta)&
=\zeta^2\int^1_0d\alpha\,g_e(\alpha)
\int^\alpha_0 d\beta\, \widehat E(\alpha\zeta,\beta\zeta,0)\,,
\end{align}
where  
\begin{eqnarray}
g_h(\alpha)=\alpha \int^1_\alpha \frac{d\tau}{\tau^2}\,f_h(\tau)\,,
\qquad
g_e(\alpha)=\int^1_\alpha d\tau\,f_e(\tau)\,.
\end{eqnarray}
These relations can be inverted to give
\begin{eqnarray}
f_h(\alpha)=g_h(\alpha)-\alpha g'_h(\alpha)+\delta(\bar \alpha)g_h(1)\,,
\qquad
f_e(\alpha)=-g'_e(\alpha)+\delta(\bar \alpha)g_e(1)\,.
\label{eq:3->2inv}
\end{eqnarray}
Thus the reduction to a two-particle form is possible for arbitrary weight functions $g(\alpha)$, but requires a very specific integration weight 
over $\beta$, i.e. over the position of the gluon field. 
Note that this weight factor can sometimes be simplified by virtue of the symmetry relations in Eqs.~\eqref{def:H-sym}, \eqref{def:E-sym}.
For example, using~\eqref{def:H-sym} for $w(\beta) = 1$ and $w(\beta) = \beta^2$ one obtains
\begin{align}
\int_0^\alpha d\beta \,\widehat H(\alpha\zeta,\beta\zeta,0)= 0\,,
&&
\int_0^\alpha d\beta \,\beta^2\, \widehat H(\alpha\zeta,\beta\zeta,0)&= \alpha \int_0^\alpha d\beta \,\beta\, \widehat H(\alpha\zeta,\beta\zeta,0)\,,  
\end{align}
etc. Similarly
\begin{align}
\int_0^\alpha d\beta \,\beta\, \widehat E(\alpha\zeta,\beta\zeta,0)&=\frac{\alpha}{2} \int_0^\alpha d\beta \, \widehat E(\alpha\zeta,\beta\zeta,0)\,.
\end{align}
 We use these relations to rewrite the quark-antiquark-gluon contributions in terms of $h_L^{\text{tw3}}$ and $e$ 
 whenever possible, as it allows one to reduce the nonperturbative input.

\subsection{Position space: Ioffe-time distributions}
\label{sec:qITDs}

The position space results are straightforward to derive from the light-ray OPE expressions 
in Appendix~\ref{app:OPE}, subtracting singular $1/\epsilon$ terms and taking the nucleon matrix elements. 
In the following expressions
\begin{eqnarray}
\mathrm{L}_z=\ln\(\frac{-z^2 \mu^2}{4\,e^{-2\gamma_E}}\),
\label{Lz}
\end{eqnarray}
and the plus-distribution is defined as usual,
\begin{eqnarray}
\int_0^1 d\alpha\, f(\alpha)\big(g(\alpha)\big)_+=\int_0^1 d\alpha\, \big(f(\alpha)-f(1)\big)g(\alpha)\,.
\end{eqnarray}

The qITD $\mathcal{H}_1$ at NLO reads
\begin{eqnarray}\label{qITD:h1}
\mathcal{H}_1(\zeta,z^2;\mu)&=&\widehat{\delta q}(\zeta;\mu)+a_s\int_0^1 d\alpha \, \mathbf{C}_1^{(1)}(\alpha,{\rm L}_z)\,\widehat{\delta q}(\alpha\zeta;\mu)\, ,
\end{eqnarray}
where
\begin{eqnarray}\label{C11}
\mathbf{C}_1^{(1)}(\alpha,{\rm L}_z)=4C_F\Big[\(-{\rm L}_z \frac{\alpha}{1-\alpha}-\frac{\alpha+2\ln\bar \alpha}{1-\alpha}\)_++\delta(\bar \alpha)({\rm L}_z+1)\Big]
\end{eqnarray}
with $C_F=(N_c^2-1)/(2N_c)$ being the quadratic Casimir operator of a general $SU(N_c)$ gauge theory.

The qITD $\mathcal{H}_L$ splits into twist-two and twist-three terms
\begin{eqnarray}\label{qITD:hL}
\mathcal{H}_L(\zeta,z^2)=\mathcal{H}^{\rm tw2}_L(\zeta,z^2)+\mathcal{H}^{\rm tw3}_L(\zeta,z^2)\, .
\end{eqnarray}
The corresponding expressions  are
\begin{align}\label{HLtw2}
{\cal H}_{L}^{\rm tw2}(\zeta,z^2)&=
\widehat{h}_{L}^{\rm tw2}(\zeta;\mu)
+a_s\int^1_0d\alpha\,{\bf C}_{1}^{(1)}(\alpha,{\rm L}_z)\,\widehat{h}_L^{\rm tw2}(\alpha\zeta;\mu)\, ,
\\
{\cal H}_{L}^{\rm tw3}(\zeta,z^2)&=
\widehat{h}_L^{\rm tw3}(\zeta;\mu)
+a_s\int^1_0d\alpha\,{\bf C}_{L, \rm 2pt}^{(1)} (\alpha,{\rm L}_z) \widehat{h}_L^{\rm tw3} (\alpha\zeta;\mu) 
+2\zeta^2a_s{\bf C}_{L,\rm 3pt}^{(1)}\otimes \widehat{H}\,,
\end{align}
where ${\bf C}_{1}^{(1)}$ is given in Eq.~(\ref{C11}),
\begin{align}
{\bf C}_{L,\rm 2pt}^{(1)}(\alpha,{\rm L}_z)&=
{\bf C}_{1}^{(1)}(\alpha,{\rm L}_z) +N_c\Big[{\rm L}_z(1+\delta(\bar \alpha))+1+2\delta(\bar \alpha)\Big]\, ,
\label{CL2}\\
{\bf C}_{L,\rm 3pt}^{(1)}\otimes \widehat{H}&=
-{\rm L}_z\, {\rm P}^{\text{tw3}}\otimes \widehat{H}
+\int_0^1 d\alpha \Big\{
2N_c\int_0^\alpha d\beta \ln\bar \beta\, \widehat{H}(\alpha\zeta,\beta \zeta,0)
\notag \\ &
\quad
+\frac{1}{N_c}\int_\alpha^1 d\beta \Big[\frac{2\bar \beta (\alpha+\ln \bar \alpha)}{1-\alpha}-\beta \bar \beta\Big]\widehat{H}(\alpha\zeta,\beta \zeta,0)\Big\}\, ,
\label{CL3}
\end{align}
and
\begin{align}\label{Ptw3}
{\rm P}^{\text{tw3}}\otimes \widehat{H}=\frac{1}{N_c}\int_0^1 d\alpha \int_\alpha^1 d\beta \frac{-\alpha \bar \beta}{1-\alpha}\widehat{H}(\alpha\zeta,\beta \zeta,0)\, .
\end{align}
Note that ${\bf C}_{L,\rm 2pt}^{(1)} (\alpha,{\rm L}_z) \slashed{=} {\bf C}_{L,\rm tw2}^{(1)} (\alpha,{\rm L}_z)$, i.e. the coefficient function of the two-particle
twist-three contributions is not equal to the twist-two coefficient function.
The term $\sim {\bf C}_{L,\rm 3pt}^{(1)}\otimes \widehat{H}$ presents a ``genuine'' three-particle contribution in the sense that it cannot be 
rewritten in terms of $\widehat{h}_L^{\rm tw3}$ using (\ref{eq:3->2inv}).
It contains a logarithmic contribution $\sim \mathrm{L}_z$ suppressed by a color factor $1/N_c$, in agreement with \cite{Balitsky:1996uh}.
but also a color-enhanced term $\propto N_c$ independent of ${\rm L}_z$.  

For the scalar case we get
\begin{align}\label{qITD:e}
{\cal E} (\zeta,z^2)&= \Big(1+6C_Fa_s\Big)\Sigma_{q} + {\cal E}_{\rm nl}(\zeta,z^2)\,,
\notag\\
{\cal E}_{\rm nl}(\zeta,z^2)&=
\widehat{e}_{\rm nl}(\zeta;\mu)
+a_s\int^1_0d\alpha\, {\bf C}_{S,\rm 2pt}^{(1)}(\alpha,{\rm L}_z) \widehat{e}_{\rm nl}(\alpha\zeta;\mu)
+2\zeta^2 a_s{\bf C}_{S,\rm 3pt}^{(1)}\otimes \widehat{E}\,,
\end{align}
where 
\begin{align}\label{CS2}
{\bf C}_{S,\rm 2pt}^{(1)}&=4C_F\left[\left(-\frac{{\rm L}_z}{\bar\alpha}
-\frac{\alpha+2\ln\bar\alpha}{1-\alpha}\right)_+ 
+\delta(\bar\alpha)\right]
+N_c\Big[-{\rm L}_z(1-\delta(\bar \alpha))-1+2\delta(\bar \alpha)\Big]\, ,
\\
{\bf C}_{S, \rm 3pt}^{(1)}\otimes \widehat{E}
&=-{\rm L}_z {\rm P}^{\rm tw3}\otimes \widehat{E}
+\int^1_0d\alpha \biggl\{ 2N_c\int^\alpha_0 d\beta\,\ln\bar\beta \,\widehat{E}(\alpha\zeta,\beta\zeta,0)
\notag\\\qquad
&
\qquad\qquad
+\frac{1}{N_c}\int_\alpha^1d\beta\,\left[\frac{2\bar\beta (\alpha+\ln\bar\alpha)}{1-\alpha}+\bar\beta\right]
\widehat{E}(\alpha\zeta, \beta \zeta,0)\biggr\}\, .
\label{CS3}
\end{align}
We have checked that the resulting scale dependence of the coefficient functions is in agreement with  
the twist-three evolution equations for (chiral-odd) three-particle distributions, see App.~\ref{app:evolution_kernel}.
Note that the evolution kernels ${\rm P}^{\rm tw3}$ \eqref{Ptw3} for $\widehat{E}$ and $\widehat{H}$ coincide as these
distributions satisfy the same evolution equation \eqref{app:kernel1}. 
The expressions for the non-logarithmic parts, ${\bf C}_{L, \rm 3pt}^{(1)}$ \eqref{CL3} and  ${\bf C}_{S, \rm 3pt}^{(1)}$ \eqref{CS3},
are also the same up to the last term, $-\beta\bar\beta$ vs. $\bar\beta$, in the square brackets.

The $\Sigma$-term in Eq.~\eqref{qITD:e} does not receive a logarithmic contribution $\sim {\mathrm L}_z$ as the one-loop scale dependence of the
quark condensate is cancelled against the scale dependence of the renormalized 
nonlocal operator \eqref{def:op-main}.
To one-loop accuracy~\cite{Shifman:1987rj} the renormalization constant reads,
\begin{align}\label{def:ZO}
Z_{\mathcal{O}}&=1-a_s\frac{3C_F}{\epsilon}.
\end{align}

\subsection{Momentum space:  pseudodistributions}
\label{sec:pPDFs}

We define pPDFs~\cite{Radyushkin:2017cyf} as Fourier transforms of qITDs with respect to the Ioffe-time $\zeta$,
\begin{eqnarray}
\label{def:pseudo1}
\mathfrak{h}_1(x,z^2)&=&\int \frac{d\zeta}{2\pi}e^{-ix\zeta}\mathcal{H}_1(\zeta,z^2),
\\\label{def:pseudoL}
\mathfrak{h}_L(x,z^2)&=&\int \frac{d\zeta}{2\pi}e^{-ix\zeta}\mathcal{H}_L(\zeta,z^2),
\\\label{def:pseudoT}
\mathfrak{e}(x,z^2)&=&\int \frac{d\zeta}{2\pi}e^{-ix\zeta}\mathcal{E}(\zeta,z^2).
\end{eqnarray}
Pseudodistributions have a natural ``partonic'' support $|x|<1$~\cite{Radyushkin:2016hsy}. 
Making the Fourier transformation of Eqs.~\eqref{qITD:h1}, \eqref{qITD:hL}, \eqref{qITD:e} we obtain
\begin{eqnarray}
\mathfrak{h}_1(x,z^2)&=&
\delta q(x)+a_s \int_{|x|}^1 \frac{d\alpha}{\alpha}\mathfrak{C}_1^{(1)}(\alpha,{\rm L}_z) \delta q\left(\frac{x}{\alpha}\right),
\\\nn
\mathfrak{h}_{L}(x,z^2)&=&
h_L(x)+a_s\int^1_{|x|}\frac{d\alpha}{\alpha}\left(\mathfrak{C}_{1}^{(1)}(\alpha,{\rm L}_z)h_L^{\rm tw2}\left(\frac{x}{\alpha}\right)+\mathfrak{C}_{L,\rm 2pt}^{(1)}(\alpha, {\rm L}_z) h_L^{\rm tw3}\left(\frac{x}{\alpha}\right)\right)
\\
&&\qquad\qquad
+2a_s\mathfrak{C}_{L,\rm 3pt}^{(1)}\otimes H\, ,
\\\nn
\mathfrak{e}(x,\zeta^2)&=&
e(x)+6C_Fa_s\delta(x)\Sigma_{\bar qq}
+a_s\int^1_{|x|}\frac{d\alpha}{\alpha} \mathfrak{C}_{S,\rm 2pt}^{(1)}(\alpha,{\rm L}_z) e_{\rm nl}\left(\frac{x}{\alpha}\right)
\\
&&\qquad\qquad
+2a_s \mathfrak{C}^{(1)}_{S, \rm 3pt}\otimes E\,,
\end{eqnarray}
where
\begin{alignat}{2}
\mathfrak{C}_1^{(1)}(\alpha,{\rm L}_z)&=\mathbf{C}_1^{(1)}(\alpha,{\rm L}_z)\, ,&\qquad  \text{Eq.\,(\ref{C11})}\, ,
\notag\\
\mathfrak{C}_{L,\rm 2pt}^{(1)}(\alpha, {\rm L}_z)&=\mathbf{C}_{L,\rm 2pt}^{(1)}(\alpha, {\rm L}_z)\, ,&\qquad \text{Eq.\,(\ref{CL2})}\, ,
\notag\\
\mathfrak{C}_{S,\rm 2pt}^{(1)}(\alpha,{\rm L}_z)&=\mathbf{C}_{S,\rm 2pt}^{(1)}(\alpha,{\rm L}_z)\, ,&\qquad \text{Eq.\,(\ref{CS2})}\, ,
\end{alignat}
and
\begin{align}
\mathfrak{C}_{L,\rm 3pt}^{(1)}\otimes H&=
-{\rm L_z}\mathfrak{P}^{\rm tw3}\otimes H
+\int [dx]\int^1_0d\alpha \Bigg\{
\nn\\\nn &
-2N_c\left(\frac{1}{1-\alpha}\right)_+\left(\frac{\delta(x-\alpha x_3)}{x_1x_3}+\frac{\delta(x+x_1+\alpha x_2)}{x_1x_2}\right)
\notag\\\nn &
+\frac{1}{N_c}
\bigg[
2\frac{\alpha+\ln\bar \alpha}{1-\alpha}\frac{\delta(x-\alpha x_3)-\delta(x+x_2+\alpha x_1)}{x_2^2}
-2\left(\frac{\alpha}{1-\alpha}\right)_+\frac{\delta(x-\alpha x_3)}{x_2x_3}
\\ &
\qquad
+(1-2\alpha)\Big(\frac{\delta(x+\alpha x_2)}{x_1x_2}+\frac{\delta(x-\alpha x_3)}{x_1x_3}\Big)
\bigg]\Bigg\}H(x_1,x_2,x_3)\, ,
\\
\mathfrak{C}_{S,\rm 3pt}^{(1)}\otimes E&=
-{\rm L_z}\mathfrak{P}^{\rm tw3}\otimes E
+\int [dx]\int^1_0d\alpha \Bigg\{
\nn\\\nn &
-2N_c\left(\frac{1}{1-\alpha}\right)_+\left(\frac{\delta(x-\alpha x_3)}{x_1x_3}+\frac{\delta(x+x_1+\alpha x_2)}{x_1x_2}\right)
\\\nn &
+\frac{1}{N_c}
\bigg[
2\frac{\alpha+\ln\bar \alpha}{1-\alpha}\frac{\delta(x-\alpha x_3)-\delta(x+x_2+\alpha x_1)}{x_2^2}
-2\left(\frac{\alpha}{1-\alpha}\right)_+\frac{\delta(x-\alpha x_3)}{x_2x_3}
\\ &\qquad
+\frac{\delta(x)}{x_2x_3}+\frac{\delta(x+\alpha x_2)}{x_1x_2}+\frac{\delta(x-\alpha x_3)}{x_1x_3}
\bigg]\Bigg\}E(x_1,x_2,x_3)\, ,
\end{align}
with
\begin{align}
\mathfrak{P}^{\rm tw3}\otimes H&=\frac{1}{N_c}\int [dx]\int^1_0d\alpha 
\bigg(
\frac{\alpha}{1-\alpha}\frac{\delta(x+x_2+\alpha x_1)-\delta(x-\alpha x_3)}{x_2^2}
\nn\\&
+\frac{\delta(x-x_3)-\delta(x-\alpha x_3)}{x_2x_3}\bigg)H(x_1,x_2,x_3)\, .
\end{align}
The integrands in these expressions are finite at $x_i\to0$ or $\alpha\to1$. 
All expressions are defined for $-1<x<1$.

\subsection{Momentum space:  quasidistributions}
\label{sec:qPDFs}

Following Refs.~\cite{Ji:2013dva,Izubuchi:2018srq} we define qPDFs as Fourier transforms of the qITDs 
with respect to the distance $z\equiv|z|$. The orientation of the vector $z^\mu$ is fixed:
\begin{eqnarray}\label{def:quasi1}
\mathtt{h}_1(x,p_v)&=&p_v\int \frac{dz}{2\pi}e^{-ixzp_v}\mathcal{H}_1(zp_v,z^2)\,,
\\\label{def:quasiL}
\mathtt{h}_L(x,p_v)&=&p_v\int \frac{dz}{2\pi}e^{-ixzp_v}\mathcal{H}_L(zp_v,z^2)\,,
\\\label{def:quasiE}
\mathtt{e}(x,p_v)&=&p_v\int \frac{dz}{2\pi}e^{-ixzp_v}\mathcal{E}(zp_v,z^2)\,,
\end{eqnarray}
where $v^\mu=z^\mu/|z|$ is the unit vector along $z^\mu$, and $p_v=(p \cdot v)$. 

The derivation of the one-loop coefficient functions for the qPDFs is more involved due to the terms containing
$\mathrm{L}_z$ \eqref{Lz},  which are also the ones responsible for extending the support property of qPDFs 
beyond the partonic region $|x|<1$ to $|x|<\infty$. 
To avoid complications of the direct Fourier transformation, we use the identity
\begin{eqnarray}\label{def:pseudo->quasi}
\mathtt{h}_1(x,p_v)=\int \frac{d\zeta}{2\pi}\int_{-1}^1 dy~e^{i(y-x)\zeta}\mathfrak{h}_1\(y,\frac{\zeta^2}{p_v^2}\), 
\end{eqnarray}
and similar for $\mathtt{h}_L$ and $\mathtt{e}$. Starting from this identity, ond can obtain
a relatively simple relation between the coefficient functions for the qPDFs and pPDFs.
Details of the derivation and explicit form of this relation can be found in App.~B of Ref.~\cite{Braun:2021aon}.

In the expressions given below we use the following notation:
\begin{align}
\mathrm{L}_p& =\ln\(\frac{\mu^2}{4 y^2 p_v^2}\),
\end{align}
and
\begin{align}
\(f(x)\)_\oplus&= f(x)-\delta(\bar x)\int_1^\infty\! f(x)\,dx\,,
\\
\(f(x)\)_\ominus&= f(x)-\delta(x)\int_{-\infty}^0\! f(x)\,dx\,.
\end{align}
We obtain
\begin{align}
\mathtt{h}_{1}(x,p_v)&=\delta q(x)+a_s\int^1_{-1}\frac{dy}{|y|}\,\mathtt{C}_1^{(1)}\Big(\frac{x}{y},{\rm L}_p\Big) \delta q(y),
\notag\\
\mathtt{h}_{L}(x,p_v)&=h_L(x)+a_s\int^1_{-1}\frac{dy}{|y|}\Big(
\mathtt{C}_1^{(1)}\Big(\frac{x}{y},{\rm L}_p\Big) h_L^{\text{tw2}}(y)
+\mathtt{C}_{L,\text{2pt}}^{(1)}\Big(\frac{x}{y},{\rm L}_p\Big) h_L^{\text{tw3}}(y)\Big)
\notag\\&\qquad +2 a_s \mathtt{C}_{L,\text{3pt}}^{(1)}\otimes H\,,
\\
\mathtt{e}(x,p_v)& = e(x)+6C_Fa_s \delta(x)\Sigma_{\bar q q}+a_s\int^1_{-1}\frac{dy}{|y|}\,
\mathtt{C}_{S,\text{2pt}}^{(1)}\Big(\frac{x}{y},{\rm L}_p\Big) e_{\text{nl}}(y)
\notag\\&\qquad +2 a_s \mathtt{C}_{S,\text{3pt}}^{(1)}\otimes E\,.
\end{align}
The two-particle coefficient functions are
\begin{eqnarray}\label{qPDF:C1}
\mathtt{C}_1^{(1)}(x,{\rm L}_p)&=&\theta(0<x<1)\mathbf{C}_1^{(1)}(x,{\rm L}_p)+4C_F\mathtt{s}_1(x)\, ,
\\
\mathtt{C}_{L,\text{2pt}}^{(1)}(x,{\rm L}_p)&=&
\theta(0<x<1)\mathbf{C}_{L,\text{2pt}}^{(1)}(x,{\rm L}_p)+4C_F\mathtt{s}_1(x)+N_c\mathtt{s}_2(x)+N_c{\rm L}_p\delta(\bar x)\, ,
\\
\mathtt{C}_{S,\text{2pt}}^{(1)}(x,{\rm L}_p)&=&
\theta(0<x<1)\mathbf{C}_{S,\text{2pt}}^{(1)}(x,{\rm L}_p)+4C_F(\mathtt{s}_1(x)-\mathtt{s}_2(x))-N_c\mathtt{s}_2(x)\, ,
\end{eqnarray}
where it is tacitly assumed that all contributions $\sim\delta(\bar x)$ (e.g., in~\eqref{C11}) are included. 

The three-particle coefficient functions are
\begin{align}
\mathtt{C}_{L,\text{3pt}}^{(1)}\otimes H&=
\theta(|x|<1)\mathfrak{C}_{L,\text{3pt}}^{(1)}\otimes H\Big|_{{\rm L}_z\to {\rm L}_p}
\nn\\&
+\frac{1}{N_c}\int [dx]\int^1_0d\alpha 
\bigg(
\mathtt{s}_1(\alpha)\frac{\delta(x+x_2+\alpha x_1)-\delta(x-\alpha x_3)}{x_2^2}
\nn\\&\qquad\qquad
-\mathtt{s}_2(\alpha)\frac{\delta(x-x_3)-\delta(x-\alpha x_3)}{x_2x_3}\bigg)H(x_1,x_2,x_3)\,,
\\
\mathtt{C}_{S,\text{3pt}}^{(1)}\otimes E&=
\theta(|x|<1)\mathfrak{C}_{S,\text{3pt}}^{(1)}\otimes E\Big|_{{\rm L}_z\to {\rm L}_p}
\nn\\&
+\frac{1}{N_c}\int [dx]\int^1_0d\alpha 
\bigg(
\mathtt{s}_1(\alpha)\frac{\delta(x+x_2+\alpha x_1)-\delta(x-\alpha x_3)}{x_2^2}
\nn\\&\qquad\qquad
-\mathtt{s}_2(\alpha)\frac{\delta(x-x_3)-\delta(x-\alpha x_3)}{x_2x_3}\bigg)E(x_1,x_2,x_3)\,.
\end{align}
In these expressions
\begin{eqnarray}
\mathtt{s}_1(\alpha)&=&\left\{\begin{array}{ll}
\Ds
\left(\frac{\alpha}{1-\alpha}\ln\left(\frac{\alpha}{-\bar \alpha}\right)-\frac{1}{1-\alpha}\right)_{\oplus}, & \alpha>1\, ,
\\\Ds
\left(\frac{\alpha \ln(\alpha\bar \alpha)}{1-\alpha}+\frac{1+2\ln\bar \alpha}{1-\alpha}\right)_{+}, & 0<\alpha<1\, ,\qquad
\\\Ds
\left(\frac{\alpha}{1-\alpha}\ln\left(\frac{\bar \alpha}{-\alpha}\right)+\frac{1}{1-\alpha}\right)_{\ominus}, & \alpha<0\, ,
\end{array}\right.
\\ 
\mathtt{s}_2(\alpha)&=&\left\{\begin{array}{ll}
\Ds
\left(-\ln\left(\frac{\alpha}{-\bar \alpha}\right)-\frac{1}{1-\alpha}\right)_{\oplus}, & \alpha>1\, ,
\\\Ds
\left(-\ln(\alpha\bar \alpha)+\frac{1}{1-\alpha}\right)_{+}, & 0<\alpha<1\, ,\qquad
\\\Ds
\left(-\ln\left(\frac{\bar \alpha}{-\alpha}\right)+\frac{1}{1-\alpha}\right)_{\ominus}, & \alpha<0\, .
\end{array}\right.
\end{eqnarray}
The integrals over delta-functions can be evaluated, revealing 30 distinct integration domains 
for the ``genuine'' three-point contribution each of which has a distinct coefficient function. 

The qPDF $\mathtt{h}_1(x,p_v)$ has previously been considered in Refs.~\cite{Liu:2018hxv,Alexandrou:2018eet}. 
Our expression for the one-loop coefficient function agrees with the result of~\cite{Liu:2018hxv}%
\footnote{We thank Jian-Hui Zhang for providing us the analytic expression for the result used in Ref.~\cite{Liu:2018hxv}, 
but not written there explicitly.}, 
but does not agree with the expression derived in~\cite{Alexandrou:2018eet}.
Our result in \eqref{qPDF:C1} reads in explicit form:
\begin{eqnarray}
\mathtt{C}_1^{(1)}(x,{\rm L}_p)=4C_F \delta(\bar x)({\rm L}_p+1)
+4C_F \left\{
\begin{array}{ll}
\Ds
\left(\frac{x}{1-x}\ln\left(\frac{x}{-\bar x}\right)-\frac{1}{\bar x}\right)_{\oplus}, & x>1\, ,
\\\Ds
\left(\frac{-x}{1-x}[{\rm L}_p-\ln(x\bar x)]+1\right)_{+}, & 0<x<1\, ,\qquad
\\\Ds
\left(\frac{x}{1-x}\ln\left(\frac{\bar x}{-x}\right)+\frac{1}{\bar x}\right)_{\ominus}, & x<0\, .
\end{array}\right.
\end{eqnarray}
Here the first term $\sim \delta(\bar x)$ depends on the normalization scheme,  see e.g.~\cite{Izubuchi:2018srq}. 
In particular, it drops out if one uses the short-distance normalization scheme.

\section{Discussion}
\label{sec:disc}

We have formulated the factorization theorem for the chiral-odd space-like correlation functions (\ref{def:ITD}) 
in terms of parton distributions to the twist-three accuracy  and calculated the corresponding coefficient functions at NLO. 
These objects can be computed by lattice QCD methods, see e.g. Refs.~\cite{Bhattacharya:2021moj, Liu:2018hxv}, 
granting access to the chiral-odd twist-three functions. 
The utility of these studies depends, however, on the possibility to subtract the twist-two contribution and reveal 
the desired twist-three part. A paradigm case is provided by the DIS structure function $g_2(x,Q^2)$ for which
the twist-two contribution can be calculated in terms of $g_1(x,Q^2)$ and subtracted from the experimental data, 
at least in principle. The situation with qPDFs and pPDFs is more complicated.

As we have demonstrated in Ref.~\cite{Braun:2021aon}, the Wandzura-Wilczek relation does not hold for the chiral-even axial-vector 
qITDs and, hence, neither for pPDFs and qPDFs. Thus the twist-two contributions cannot be subtracted exactly, but only up to
a certain order in perturbation theory. 
The situation for the chiral-odd distributions turns out to be more encouraging.

The distribution $\mathcal{H}_L$ includes a twist-two contribution \eqref{HLtw2}. However, at one loop it involves the {\it same} 
coefficient function $\mathbf{C}_1^{(1)}(\alpha,{\rm L}_z)$ \eqref{C11} as for $\mathcal{H}_1$.
This equality is not accidental and is a consequence of the fact that in the chiral-odd case no additional tensor structure can arise 
from loop integrals as  $\mathcal{H}_L^{\text{tw2}}$ and $\mathcal{H}_1$ are just different Lorentz projections of the same twist-two operator
with an open vector index. This is in contrast to the chiral-even case where terms $\sim z^\mu z^\nu/z^2$ (in addition to $g^{\mu\nu}$ tensor that is present already at the tree-level) emerge via loop-integral giving rise to additional
contributions~\cite{Braun:2021aon}. This argument applies to all orders in perturbation theory, so that the equality 
of the  coefficient functions for $\mathcal{H}_L^{\text{tw2}}$ and $\mathcal{H}_1$ is \textit{exact}. As a consequence, the twist-two
contribution to ${\cal H}_L$ in position space distributions can be eliminated by the analogue of the Jaffe-Ji relation~\cite{Jaffe:1991kp} 
\begin{eqnarray}\label{discussion1}
\mathcal{H}^{\text{tw3}}_L(\zeta,z^2)=\mathcal{H}_L(\zeta,z^2)-2\int_0^1 d\alpha\,\alpha \,\mathcal{H}_1(\alpha\zeta,z^2)\,,
\end{eqnarray}
cf. \eqref{hL-tw2}. The same subtraction can be implemented in pPDFs (here $x>0$):
\begin{eqnarray}\label{discussion2}
\mathfrak{h}^{\text{tw3}}_L(x,z^2)=\mathfrak{h}_L(x,z^2)-2x\int_x^1 \frac{dy}{y^2}\,\mathfrak{h}_1(y,z^2)\,,
\end{eqnarray}
but cannot be applied directly to qPDFs. The reason is that in this case the Fourier transformation is applied to the complete $z$-dependence
without distinguishing whether it originates from $\zeta = (p\cdot z)$ or $z^2$.
As the result, the relevant analogue of Eq.~\eqref{discussion1} involves $z$-dependence in the logarithms ${\rm L}_z$
\begin{eqnarray}
&&\mathcal{H}_L(\zeta,z^2)-2\int_0^1 d\alpha\,\alpha\, \mathcal{H}_1(\alpha\zeta,\alpha^2 z^2)
\\\nn &&\qquad=
\mathcal{H}_L^{\text{tw3}}(\zeta,z^2)+
8a_sC_F\int_0^1 d\alpha\, \alpha\Big(
2 \ln\alpha\ln\bar\alpha-\ln^2\alpha +2\Li_2(\bar \alpha)
\Big)\widehat{\delta q}(\alpha \zeta)+\mathcal{O}(a_s^2)\, ,
\end{eqnarray}
and after the Fourier transformation 
\begin{eqnarray}
&&\mathtt{h}_L(x,p_v)-2\int_{|x|}^1 dy\, \mathtt{h}_1\left(\frac{x}{y},p_v\right)
\\\nn &&\qquad=
\mathtt{h}_L^{\text{tw3}}(x,p_v)+
8a_sC_F\int_{|x|}^1 dy\Big(
2\ln y\ln\bar y-\ln^2 y+2\Li_2(\bar y)
\Big)\delta q\left(\frac{x}{y}\right)+\mathcal{O}(a_s^2).
\end{eqnarray}
In principle, the twist-two remainder on the r.h.s. of this relation can be subtracted up to $\mathcal{O}(a_s^2)$ 
by promoting $\delta q(x)\to \mathtt{h}_1(x)$.
 
As well known, the extraction of the momentum-fraction dependence of the PDFs from lattice calculations requires hadron 
sources with large momentum, which is a notorious problem. Large momenta are not necessary, however, to access an overall normalization of the twist-three
contributions encoded in the matrix elements of the local twist-three operators of the lowest dimension. To this end the $\zeta\to 0$ expansion 
of the qITDs is sufficient. One obtains
\begin{align}
  \mathcal{H}^{\text{tw3}}_L(\zeta,z^2) & = -\frac{i \zeta^3  h_3}{6} 
\biggl\{1+a_s\left[{\rm L}_z\left(\frac{65N_c}{12}-\frac{25}{12N_c}\right)-\frac{289N_c}{36}+\frac{29}{36N_c}\right]\biggr\} +\mathcal{O}(\zeta^4,z^2)\,,   
\notag\\
\mathcal{E}(\zeta,z^2)&=(1+6 C_F a_s)\Sigma_q
\notag\\ &\quad
-\frac{\zeta^2 e_2}{2}\biggl\{
1+a_s\Big[{\rm L}_z
\Big(\frac{11}{3}N_c-\frac{8}{3 N_c}\Big)
-\frac{11N_c}{3}+\frac{11}{3N_c}\Big]\biggr\}+\mathcal{O}(\zeta^3,z^2)\,,
\end{align}  
where 
\begin{eqnarray}
\langle p,s|\bar q \sigma^{\mu n}\gamma_5 \big[{iD}_n, g F_{\mu n }\big] q  |p,s\rangle 
&=& \frac{4}{5} (s\cdot n) (p\cdot n)^2 M^2 \, h_3\, ,
\notag\\
\langle p,s|\bar q \,\sigma^{\mu n} \, g F_{\mu n }q  |p,s\rangle &=& -4 (p\cdot n)^2 M \, e_2\,,
\end{eqnarray}
with all fields assumed to be at the origin and $n^2=0$.
Equivalently
\begin{eqnarray}
h_3&=&\int_{-1}^1 dx \,x^3\, h_L^{\text{tw3}}(x)=\frac{1}{5}\int [dx]\,x_2\, H(x_1,x_2,x_3)\, ,
\\
e_2&=&\int_{-1}^1 dx \, x^2 \, e(x)=-\int [dx]\,E(x_1,x_2,x_3)\, .
\end{eqnarray}
Note that $\int [dx]\, H(x_1,x_2,x_3)=0$ as follows from the symmerty \eqref{def:H-sym}. As a consequence
$\mathcal{H}^{\text{tw3}}_L(\zeta,z^2) = \mathcal{O}(\zeta^3)$. 

To conclude, in this work we have presented, for the first time, the complete NLO analysis of the chiral-odd quasi- and pseudo-distributions of the nucleon 
to the twist-three accuracy. The results are encouraging so that we expect that the chiral-odd twist-three PDFs can be constrained from
lattice calculations in the near future.

\acknowledgments

This study was supported by Deutsche Forschungsgemeinschaft (DFG) through the Research 
Unit FOR 2926, ``Next Generation pQCD for Hadron Structure: Preparing for the EIC'', project number 40824754.
Y.J. also acknowledges the support of DFG grant SFB TRR 257.

\appendix

\section*{Appendices}

\section{Evolution kernel for twist-3 distributions}
\label{app:evolution_kernel}

The evolution equations for twist-three quark-antiquark-gluon distributions can be found in Ref.~\cite{Braun:2009mi,Ji:2014eta}. 
For readers' convenience, we collect the relevant expressions in this appendix. 

The evolution equation for the quark-antiquark-gluon correlation functions in position space has the form 
\begin{eqnarray}
\mu^2\frac{d}{d\mu^2}\widehat{H}(z_1,z_2,z_3)=-a_s[\mathbb{H}\otimes \widehat{H}](z_1,z_2,z_3),
\end{eqnarray}
where $\mathbb{H}$ is an integral operator (evolution kernel) which is the same for $\widehat{H}$ and $\widehat{E}$:
\begin{align}\label{app:kernel1}
&[\mathbb{H}\otimes \widehat{F}](z_1,z_2,z_3)=
\notag\\&=
N_c\int_0^1 d\alpha\Big[
\frac{4}{\alpha}\widehat{F}(z_1,z_2,z_3)
-\frac{\bar \alpha}{\alpha} \widehat{F}(z_{12}^\alpha,z_2,z_3)
-\frac{\bar \alpha}{\alpha} \widehat{F}(z_1,z_2,z_{32}^\alpha)
\notag\\&\qquad
-\frac{\bar \alpha^2}{\alpha} \widehat{F}(z_1,z_{21}^\alpha,z_3)
-\frac{\bar \alpha^2}{\alpha} \widehat{F}(z_1,z_{23}^\alpha,z_3)
-2\int_0^{\bar \alpha}d\beta\,\bar \beta 
\Big(
\widehat{F}(z_{12}^\alpha,z_{21}^\beta ,z_3)+
\widehat{F}(z_1,z_{23}^\beta ,z_{32}^\alpha)
\Big)\Big]
\notag\\&\qquad
-\frac{1}{N_c}\int_0^1 d\alpha \Big[
\frac{2}{\alpha}\widehat{F}(z_1,z_2,z_3)
-\frac{\bar \alpha}{\alpha} \widehat{F}(z_{13}^\alpha,z_2,z_3)
-\frac{\bar \alpha}{\alpha} \widehat{F}(z_1,z_2,z_{31}^\alpha)
\notag\\&\qquad
+2\int_{\bar \alpha}^1d\beta\,\bar \beta 
\Big(
\widehat{F}(z_{12}^\alpha,z_{21}^\beta ,z_3)+
\widehat{F}(z_1,z_{23}^\beta ,z_{32}^\alpha)
\Big)\Big]
-3C_F\widehat{F}(z_1,z_2,z_3)\, .
\end{align}
Here $\widehat{F} = \{\widehat{H},\widehat{E}\}$ and
\begin{eqnarray}
\bar \alpha =1-\alpha\, ,\qquad z_{ij}^\alpha=z_i\bar \alpha+z_j \alpha\, .
\end{eqnarray}
The last contribution, proportional to $3C_F$, is universal to all distributions and cancels against the renormalization scale dependence 
of the qPDF operator.

The corresponding expressions for distributions in momentum space are lengthy as the evolution kernels 
in different sectors $x_i\lessgtr 0$ are not the same. Explicit expressions can be found in \cite{Ji:2014eta}. 
Note that the evolution kernel \eqref{app:kernel1} becomes distinct for $\widehat{H}$ and $\widehat{E}$
once the symmetry relations \eqref{def:H-sym}, \eqref{def:E-sym} are applied.

The twist-three quark-antiquark distributions $\widehat{h}_L^{\rm tw3}$ and $\widehat{e}_{\text{nl}}$ are given by the integrals 
\eqref{tree:h-position}, \eqref{tree:e-position}. Application of the operator $\mathbb{H}$ to these expressions yields
\begin{eqnarray}\nn
&&\int_0^1 d\alpha \int_\alpha^1 d\beta \,\alpha(2\beta-\alpha)\, [\mathbb{H}\otimes \widehat{H}](\alpha\zeta,\beta \zeta,0)
=\frac{1}{N_c}\int_0^1 d\alpha\int_\alpha^1 d\beta \frac{2\alpha\bar \beta}{1-\alpha}\widehat{H}(\alpha\zeta,\beta \zeta,0)
\\ &&\qquad\qquad
+\int_0^1 d\alpha \int^\alpha_0 d\beta\, \alpha(2\beta-\alpha)\Big[
\frac{N_c}{\alpha}-4C_F\ln\Big(\frac{\bar \alpha}{\alpha}\Big)-3C_F\Big]\widehat{H}(\alpha\zeta,\beta \zeta,0)\,,
\\\nn
&&\int_0^1 d\alpha \int_\alpha^1 d\beta \,[\mathbb{H}\otimes \widehat{E}](\alpha\zeta,\beta \zeta,0)
=\frac{1}{N_c}\int_0^1 d\alpha\int_\alpha^1 d\beta \frac{2\alpha\bar \beta}{1-\alpha}\widehat{E}(\alpha\zeta,\beta \zeta,0)
\\ &&\qquad\qquad
+\int_0^1 d\alpha \int^\alpha_0 d\beta\Big[
-4C_F\ln \bar \alpha+N_c \alpha -3C_F\Big]\widehat{E}(\alpha\zeta,\beta \zeta,0)\, .
\end{eqnarray}
Here, we have used relations \eqref{def:H-sym}, \eqref{def:E-sym} to simplify the result. 
In both expressions, the contributions in the second line the gluon field is positioned in between the quark and the antiquark.
These terms can be rewritten  as a convolution of two-particle kernels with $\widehat{h}_L$ and $\widehat{e}_{\text{nl}}$. 
The remaining terms $\sim 1/N_c$ contain the gluon field outside of the quark-antiquark pair. They cannot be simplified in this way 
and contribute to the ``genuine'' three-particle part.


\section{Light-ray OPE at twist three}
\label{app:OPE}

\begin{figure}[t]
\begin{center}
 \includegraphics[width= 0.9\textwidth]{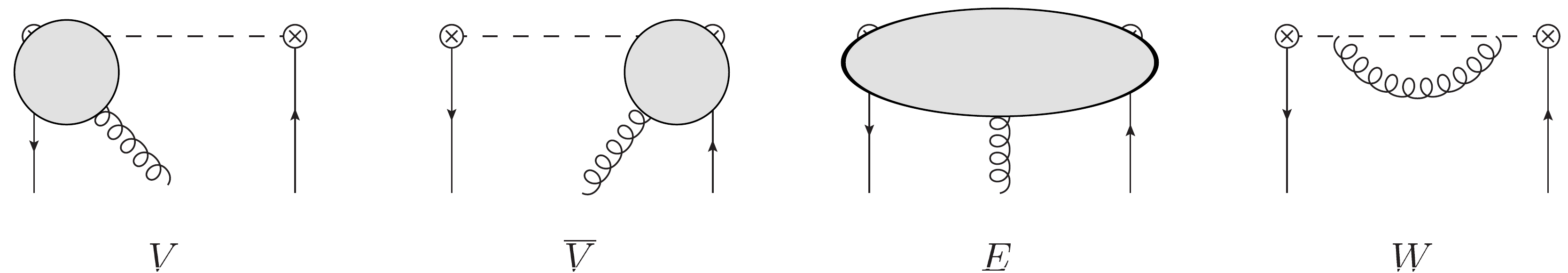}~
\caption{Contributions of different topology to the light-ray OPE of the
quark-antiquark operator}
\label{fig:Diagrams}
\end{center}
\end{figure}
%


One-loop contributions to the light-cone expansion of the nonlocal operators into 
Eq.~\eqref{def:op-main} can be divided in several classes:
Vertex corrections $V$ and $\overline{V}$, ``exchange'' 
diagrams $E$ with a hard gluon connecting the quark and the antiquark, and 
the self-energy correction $W$ to the Wilson line as shown schematically 
in Fig.~\ref{fig:Diagrams}. Each contribution is separately gauge-invariant. 
In this appendix we collect the corresponding 
expressions for twist-three contributions in $d=4-2\epsilon$ dimensions, i.e. before the subtraction 
of collinear singularities.  
We write the result in the form
\begin{align}
 \mathcal{O}^{\scriptsize\II} &=  \mathcal{O}^{\scriptsize\II}_{\text{tree}} + 
 \mathcal{N}\biggl\{ V_e + \overline{V}_e + E_e + W_e\biggr\}, 
\notag\\
 \mathcal{O}^{i\sigma_{pz}\gamma_5} &=  \mathcal{O}^{i\sigma_{pz}\gamma_5}_{\text{tree}} + 
 \mathcal{N}\biggl\{ V_h + \overline{V}_h + E_h + W_h\biggr\}, 
\end{align}
where the tree-level expressions are given in Eq.~\eqref{EOM},
and 
\begin{align}
   \mathcal N = \frac{g^2}{16\pi^\dhalf}\frac{\Gamma(\dhalf-2)}{[- z^2\mu^2]^{\dhalf-2}}\, . 
\end{align}
We obtain
\begin{align}
V_h&= 
- C_F
\bar q(z) i\sigma_{pz}\gamma_5  q(0)
+ C_F \frac{d-2}{d-3}\int_0^1\!d\alpha\, [ \alpha^{3-d}- 1]  [\bar q(z) i\sigma_{pz}\gamma_5 q(\alpha z)]_+
\notag\\&\quad
- \frac{N_c}{2}\,(pz)
\int_0^1\!d\alpha \int_{\alpha}^1\! d\beta\, \beta\frac{1-\beta^{3-d}}{3-d}
\bar q(z) g F_{\mu z } (\beta z) \sigma^{\mu z}\gamma_5 q(\alpha z) 
\notag\\&\quad
+ (C_F-\tfrac12 N_c) (pz)
 \int_0^1 \!d\alpha\,\int_0^\alpha d \beta\,\beta\, \frac{1-\alpha^{3-d}}{3-d}\, 
\bar q(z) g F_{\mu z } (\beta z) \sigma^{\mu z}\gamma_5 q(\alpha z) 
\notag\\
 \overline{V}_h&=
- C_F 
\bar q(z) i\sigma_{pz}\gamma_5  q(0)
+ C_F  \frac{d-2}{d-3}\int_0^1\!  [ \alpha^{3-d}- 1]  [\bar q(\bar\alpha z) i\sigma_{pz}\gamma_5 q(0)]_+
\notag\\&\quad
+ \frac{N_c}{2} (pz)  
\int_0^1\!d\alpha\, \int_\alpha^{1} \! d\beta\, 
\beta \frac{1- \beta^{3-d}}{3-d} 
\bar q(\bar \alpha z) 
  g F_{\mu z}\big(\bar \beta  z\big)\sigma^{\mu z} \gamma_5 q(0)  
\notag\\&\quad
- \left(C_F\!-\!\tfrac12 N_c\right)\,(pz) 
\int_0^{1} \!d\alpha \frac{1-\alpha^{3-d}}{3-d} \int_0^{\alpha}\! d\beta\,\beta\,    
\bar q(\bar\alpha z)
          g F_{\mu z}\big(\bar \beta  z\big)\sigma^{\mu z} \gamma_5 q(0) 
\end{align}

\begin{align}
V_e&= 
- C_F 
\bar q(z) q(0)
+ C_F  \frac{d-2}{d-3}\int_0^1\!  [ \alpha^{3-d}- 1]  [\bar q(z) q(\alpha z)]_+
\notag\\&\quad
- \frac{N_c}{2}\, 
\int_0^1\!d\alpha \int_{\alpha}^1\! d\beta\, \beta\frac{1-\beta^{3-d}}{3-d}
\bar q(z) g F_{\mu z } (\beta z) \sigma^{\mu z} q(\alpha z) 
\notag\\&\quad
+ (C_F-\tfrac12 N_c) 
 \int_0^1 \!d\alpha\,\int_0^\alpha d \beta\,\beta\, \frac{1-\alpha^{3-d}}{3-d}\, 
\bar q(z) g F_{\mu z } (\beta z) \sigma^{\mu z} q(\alpha z) 
\notag\\
 \overline{V}_e&=
- C_F  \bar q(z) q(0)
+ C_F  \frac{d-2}{d-3}\int_0^1\!d\alpha\,  [ \alpha^{3-d}- 1]  [\bar q(\bar\alpha z) q(0)]_+
\notag\\&\quad
- \frac{N_c}{2}   
\int_0^1\!d\alpha\, \int_\alpha^{1} \! d\beta\, 
\beta \frac{1- \beta^{3-d}}{3-d} 
\bar q(\bar \alpha z) 
  g F_{\mu z}\big(\bar \beta  z\big)\sigma^{\mu z}  q(0)  
\notag\\&\quad
+ \left(C_F\!-\!\tfrac12 N_c\right)\, 
\int_0^{1} \!d\alpha \frac{1-\alpha^{3-d}}{3-d} \int_0^{\alpha}\! d\beta\,\beta\,    
\bar q(\bar\alpha z)  g F_{\mu z}\big(\bar \beta  z\big)\sigma^{\mu z}  q(0) 
\end{align}
where the plus distribution is defined as
\begin{align}
  [\bar q(z) \Gamma q(\alpha z)]_+ &=  \bar q(z)\Gamma q(\alpha z) - \bar q(z)\Gamma q(0)\,,   
\end{align}
etc. The two-particle contributions in these expressions can be rewritten in terms of the three-particle operators
using equations of motion~\eqref{EOM}.

Further,
\begin{align}
 E_h&=  C_F (d-4)^2 \int_0^1\!dv\, \int_0^v\!du \, \bar q(vz) i\sigma_{pz}\gamma_5 q(uz)
\notag\\&\quad+
\frac12 C_F (pz) (d-4) 
\int_0^1dv\!\int_0^v\!du\! \int_u^v\! dt\,(2t-1)\, \bar q(vz) \sigma^{\mu z} g F_{\mu z}\big(t z \big)\gamma_5 q(uz)
\notag\\&\quad-
\frac{1}{2} \left(C_F\!-\!\tfrac12 N_c\right) (pz)(d-4)
\int_0^1\!dv\,\int_0^v\!du\,\int_0^1 dt \, (2t-1)\bar q(vz) \sigma^{\mu z } g F_{\mu z} (t z)\gamma_5 q(uz) 
\notag\\
 E_e&= 
 C_F d\, \bar q(0) q(0) 
+ \frac{d}{2} C_F \int_0^1\!dv\, \int_0^v\!du \!\int_u^v\!dt\,
  \bar q(vz) \sigma^{\mu z} g F_{\mu z}\big(t z \big)q(uz)  
\notag\\&\quad
- \frac12 \left(C_F\!-\!\tfrac12 N_c\right) (d-4) \int_0^1\!dv\, \int_0^v\!du \!\int_0^1\!dt\,
\bar q(vz) \sigma^{\mu z} g F_{\mu z}\big(t z \big)q(uz)  
\end{align}
and 
\begin{align}
 W &=    \frac{2 C_F}{3-d}\, \mathcal{N} \,  \bar q(x)\, \Gamma\, q(0) 
\end{align}
with $\Gamma = \{1, i\sigma_{pz}\gamma_5\}$ for $W_e$ and $W_h$, respectively.

\bibliography{bibFILE}

\bibliographystyle{JHEP}


\end{document}